\documentclass{pnastwo}

\usepackage[dvips]{graphicx}
\usepackage{amssymb,amsfonts,amsmath,color}

\begin{document}

\title{Meso-scale turbulence in living fluids}

\author{Henricus H. Wensink\affil{1}{Institute for Theoretical Physics II: Soft
Matter, Heinrich-Heine-Universit\"{a}t D\"{u}sseldorf,
Universit\"{a}tsstra{\ss}e 1, D-40225, D\"{u}sseldorf, Germany}\affil{2}{Laboratoire de Physique des Solides, Universit\'{e} Paris-Sud 11, B\^{a}timent 510, 91405 Orsay Cedex, France}, 
J\"orn Dunkel\affil{3}{DAMTP, Centre for
Mathematical Sciences, University of Cambridge, Wilberforce Road, Cambridge CB3
0WA, UK}, 
Sebastian Heidenreich\affil{4}{Physikalisch-Technische Bundesanstalt,
Abbestr. 2-12, 10587 Berlin, Germany},  
Knut Drescher\affil{3}{DAMTP, Centre for
Mathematical Sciences, University of Cambridge, Wilberforce Road, Cambridge CB3
0WA, UK}\affil{5}{Departments of Molecular Biology and Mechanical and Aerospace Engineering, Princeton University, Princeton, New Jersey 08544, USA},
Raymond E. Goldstein\affil{3}{DAMTP, Centre for
Mathematical Sciences, University of Cambridge, Wilberforce Road, Cambridge CB3
0WA, UK}, 
Hartmut L\"owen\affil{1}{Institute for Theoretical
Physics: Soft Matter, Heinrich-Heine-Universit\"at-D\"usseldorf,
Universit\"atsstra{\ss}e 1, D-40225, D\"usseldorf, Germany} and Julia M.
Yeomans\affil{6}{Rudolf Peierls Centre for Theoretical Physics, University of
Oxford, 1 Keble Road, Oxford  OX1 3NP, UK}}

\contributor{Proceedings of the National Academy of Sciences of the United States of America (2012), doi: 10.1073/pnas.1202032109}

\maketitle

\begin{article}

% %%%%%%%%%%%%%%%%%%%%%%%%%%%%%%%%
\begin{abstract}
Turbulence is ubiquitous, from oceanic currents to small-scale biological and quantum systems. Self-sustained turbulent  motion in microbial suspensions presents an intriguing example of collective dynamical behavior amongst the simplest forms of life, and is important for fluid mixing and molecular transport on the microscale. The mathematical characterization of  turbulence phenomena in active non-equilibrium fluids proves even more difficult than for conventional liquids or gases. It is not known  which features of turbulent phases in living matter are universal or system-specific, or which generalizations of the Navier-Stokes equations are able to describe them adequately. 
Here, we combine experiments, particle simulations, and continuum theory to identify the statistical properties of self-sustained meso-scale turbulence in active systems. To study how  dimensionality and boundary conditions affect collective bacterial dynamics, we measured  energy spectra and structure functions  in dense \textit{Bacillus subtilis} suspensions in quasi-2D and 3D geometries. Our experimental results for the bacterial flow statistics agree well with predictions from a minimal model for self-propelled rods, suggesting that at high concentrations the collective motion of the bacteria is dominated by short-range  interactions.  To provide a basis for future theoretical studies, we propose a minimal continuum model for incompressible bacterial flow.  A detailed numerical analysis of the 2D case shows that this theory can reproduce many of the experimentally observed features of self-sustained active turbulence.  
\end{abstract}
% %%%%%%%%%%%%%%%%%%%%%%%%%%%%%%%%

%%% PNAS rule: keywords must not appear in title & abstract

\abbreviations{
SPR, self-propelled rod; Re, Reynolds number 
%NS, Navier-Stokes
 }

%%% General introduction & importance
\dropcap{S}imple forms of life, like amoebae or bacteria, self-organize into remarkable macroscopic
patterns~\cite{2009CoWe,2011KochSub}, ranging from extended networks~\cite{2010Tero,2009XaMaFo} to complex vortices~\cite{1997Kessler,2000Be,2004DoEtAl,2007SoEtAl,2005Riedel_Science,2005Tuval_PNAS}  and swarms~\cite{2010Kearns}. These structures often bear a striking resemblance to assemblies of higher organisms (e.g.,  flocks of birds~\cite{2009Parisi} or schools of fish~\cite{2011Couzin,2006Sumpter}), and present important biological model systems to study non-equilibrium phases and their transitions~\cite{2005ToTuRa,2010Ramaswamy,1998TonerTu_PRE}. A  particularly interesting manifestation of collective behavior in microbial suspensions is the emergence of meso-scale turbulent motion~\cite{2004DoEtAl,2007SoEtAl,2007Cisneros,2008Wolgemuth}. Driven by the microorganisms' self-propulsion and their mutual interactions, such self-sustained \lq active turbulence\rq\space  can have profound effects on nutrient mixing and molecular transport in microbiological  systems~\cite{2011KochSub,2011Stocker_Review,2011ZaDuYe,2011Gollub_PNAS}. However, in spite of recent progress~\cite{2008Wolgemuth,2007Ar,2011Japan,2011Cisneros_PRE}, the phenomenology of  turbulent bacterial dynamics  is scarcely understood, and a commonly accepted theoretical description is lacking~\cite{2011KochSub,2010Ramaswamy,2000CzirokVicsek}.  The latter fact may not be surprising given that a comprehensive mathematical characterization of  turbulence in conventional fluids has remained elusive after more than a century of intense research~\cite{2004Frisch}. 

\par % open problems
In view of the  various physical and chemical pathways through which bacteria may communicate~\cite{2009CoWe,2010Kearns,2005Bassler_Review}, a basic yet unsolved problem is to identify those interactions that are responsible for the emergence of collective behavior in dense suspensions~\cite{2011KochSub,2009SoAr,2011DrescherEtAl}. Answering this question is essential for understanding whether physical mechanisms such as flagellar bundling or hydrodynamic long-range interactions are relevant to collective bacterial motion; it is also crucial for constraining the vast  number of  theoretical models that have been proposed during the past two decades~\cite{2011KochSub,2010Ramaswamy,2008Wolgemuth,2009BaMa_PNAS,2008BaMa}, but have yet to be tested  against experiments. An equally important, unresolved  issue pertains to the \lq universality\rq\space of turbulent phenomena in active systems and their relation to turbulence in passive fluids~\cite{2004Frisch}. In ordinary liquids and gases, such as water or air, turbulent vortices form due to external forcing if the Reynolds number, the ratio of inertial to viscous forces,  is very large (Re $\gg 1$). By contrast, bacteria provide an internal microscopic forcing and operate at  Re~$\sim10^{-5}$~\cite{1977Pu}. It is therefore unclear how, or to what extent, the characteristics of self-sustained turbulent states in microbial suspensions  differ from  those of classical turbulence in passive fluids. 

\par  % this work 
Here, we combine  numerical simulations, high-speed microscopic imaging  and continuum theory to identify generic statistical properties of active turbulent motion in dense bacterial systems, using  \textit{Bacillus subtilis} as a model organism.  Unlike previous investigations of collective bacterial swimming in 2D free-standing films~\cite{2007SoEtAl} and 3D bulk suspensions with liquid-gas interfaces~\cite{2004DoEtAl,2011Japan,2011Cisneros_PRE}, we conducted experiments  in closed  quasi-2D and 3D microfluidic chambers to minimize external influences and to compare the effects of boundary conditions and dimensionality. Our analysis focusses on traditional turbulence measures, such as energy spectra and velocity structure functions~\cite{2004Frisch,2002Kellay,2000Danilov}. These quantities have been widely studied for turbulent high-Re Navier-Stokes flow~\cite{2004Frisch,2002Gotoh_PhysFluids,1997Noullez_JFM,1997Camussi,1980Kraichnan, 1999LewisSwinney_PRE,1984Anselmet_JFM}, but their characteristics are  largely unknown for active fluids.  We compare our experimental results with large-scale simulations of a 2D minimal model for self-propelled rods (SPRs). In the past, similar models~\cite{2010Ginelli} have proven useful for identifying generic aspects of flocking and swarming in active systems~\cite{2010Berg_BiophysJ,Swinney_bactclust}. We find that, although the SPR model neglects details of bacterial cell-cell interactions, it is able to reproduce many features of our experimental data, suggesting that collective bacterial dynamics in dense suspensions is dominated by short-range interactions~\cite{2011DrescherEtAl}. We complement our experiments and particle-based simulation studies by identifying a minimal  continuum model for  incompressible active flow that combines elements from the Toner-Tu~\cite{2005ToTuRa,2010Ramaswamy,1998TonerTu_PRE} and  Swift-Hohenberg~\cite{1977SwiftHohenberg} theories.  

%%%%%%%%%%%%%%%%%%%%%%%%% 
\section{Theory and Experimental Results}
%%%%%%%%%%%%%%%%%%%%%%%%%

\subsection{Motivation for the SPR Model}
Self-motile bacteria may form meso-scale vortex patterns  if their concentration is sufficiently large~\cite{2004DoEtAl,2007SoEtAl,2007Cisneros,2008Wolgemuth}. At very high volume fractions ($\phi\gtrsim 40\%)$, steric repulsion and other short-range interactions (e.g., lubrication forces, flagellar  bundling of neighboring cells) can be expected to govern physical reorientation and alignment, whereas intrinsic Brownian motion effects~\cite{2011DrescherEtAl}   become less important in this collision-dominated high-density regime~\cite{2009Hoefling}.   Chemotaxis~\cite{2004DoEtAl,2007Cisneros} can strongly affect bacterial dynamics in droplets or near liquid-gas interfaces, but is less relevant in closed chambers as considered in our experiments.  Recent direct measurements of individual {\it Escherichia coli} flow fields~\cite{2011DrescherEtAl} suggest that hydrodynamic far-field interactions are   negligible for bacterial reorientation, especially when bacteria swim close to a no-slip surface. Earlier experiments~ \cite{2007SoEtAl, 2011Japan,2011Cisneros_PRE}  on 2D films and 3D bulk suspensions also show that the average swimming speeds of individual bacteria (typically of the order of $10\;\mu$m/s in isolation~\cite{2007SoEtAl,2011DrescherEtAl}) can be enhanced up to five times through collective hydrodynamic near-field effects. In the simplest approximation, however, a sufficiently dense bacterial suspension can be viewed as a system of deterministic, self-propelled, rod-like particles with an effective swimming speed~$V$ (for {\it B. subtilis} at  $\phi\sim 40\%$ we find $V\sim 30$ to $100\,\mu$m/s depending on oxygen concentration and boundary conditions). One of our objectives is to test such a minimal model  against experiments in the limit of highly concentrated suspensions and to provide systematic guidance for more accurate future models.

\subsection{Non-Equilibrium Phase Diagram of the SPR Model}
To identify generic requirements for the formation of turbulent phases in active systems, 
we performed simulations of a minimal 2D SPR model  with periodic boundary conditions (see SI Appendix for details).  In its simplest form, the model assumes that a rod-shaped self-propelled particle moves deterministically in the overdamped low-Re regime with an effective swimming speed~$V$, while interacting with the other particles by steric forces. Mutual repulsion is implemented by discreti\-zing each rod into spherical segments and imposing a repulsive Yukawa force potential $\sim\exp(-r/\lambda)/r$, where $r$ is the distance, between the segments of any two rods (i.e., the decay length $\lambda>0$  defines the effective diameter of a rod of length $\ell$). If two sufficiently long rods perform a pair-collision, this short-range interaction results in an effective nematic (apolar) alignment, before the rods become pushed apart by the repulsive force.

\par
Depending on the  effective volume filling fraction $\phi$ and the rod aspect ratio~$a$, both defined in terms of the scale parameter $\lambda$ and rod length $\ell$, the SPR model exhibits a range of qualitatively different dynamical phases (Fig.~\ref{nephase}). The numerically estimated non-equilibrium phase diagram (Fig.~\ref{nephase}A) illustrates the importance of  the effective particle \lq shape\rq\space in 2D:   Upon increasing~$\phi$, short rods undergo a transition from a dilute state~(D), with little or no cooperative motion, to a jammed state (J); this transition can be identified by the 
mean square displacement per particle, which drops off nearly two orders in magnitude along the transition curve. By contrast, very long rods ($a > 13$)  do not jam at moderate filling fractions but exhibit swarming (S) behavior and large spatiotemporal density fluctuations. Generally, the transitions from the dilute phase (D) to cooperative motion (regions S, B and T) can be characterized by the Onsager overlap density~\cite{onsager}.  Upon  increasing  $\phi$ further, very long rods tend to assemble in homogeneous lanes~(L), corresponding to quasi-smectic regions of local polar order; the swarming-to-laning transition is signaled by a discontinuous increase in the correlation length of the two-particle velocity correlation function.  The swarming (S) and laning (L) phases adjoin a so-called active bionematic~\cite{2007Cisneros} phase (B), where vortices and extended jet-like structures coexist \cite{2009Swinney_EPL,2011Cisneros_PRE}; this phase is characterized  by large fluctuations of the local vortex density.  Most importantly for the present study, however, the SPR model predicts homogeneous turbulent  states  (T) at high filling fractions and  intermediate aspect ratios $3 \lesssim a \lesssim 13$, a range that covers  typical  bacterial values (e.g., $2\lesssim a\lesssim 4$  for {\it E. coli}  and $2\lesssim a\lesssim 10$ for {\it B. subtilis} (SI Appendix, Fig. S7). The transition between bionematic and turbulent phase is also signaled by the velocity distribution,  correlation functions  and density fluctuations (SI Appendix, Fig.~S3, S4).

\subsection{Homogeneous Turbulent Phase in the SPR Model}

\begin{figure}[t]
\centering 
\includegraphics[clip=,width= 0.8 \columnwidth]{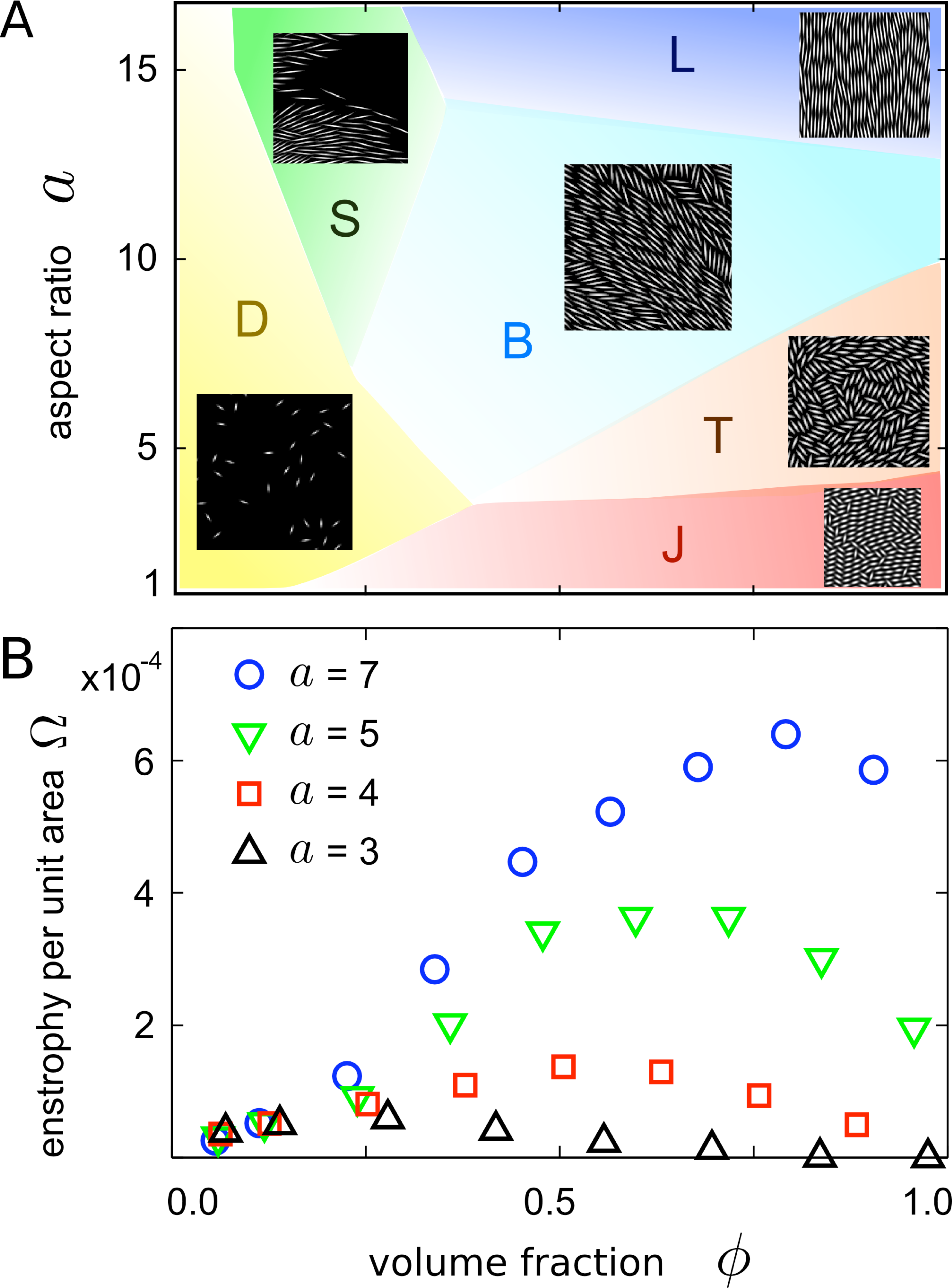} 
\caption{
(A) Schematic non-equilibrium phase diagram of the 2D SPR model and snapshots of six distinct phases from simulations:  D-dilute state,  J-jamming,  S-swarming, B-bionematic phase, T-turbulence,  L-laning (see also SI Appendix, Fig.~S2 and Movies~S01-S06). Our analysis focusses on the turbulent regime T.
\label{nephase}
(B) Enstrophy  per unit area $\Omega $ in units $(V/\lambda)^2$ for different aspect ratios $a=\ell/\lambda$, 
obtained from SPR simulations with $N\sim10^4$ to $10^5$ particles.  The maxima of the enstrophy  indicate the optimal filling fraction for 
active turbulence and mixing at a given value of the aspect ratio~$a$. Note that values $\phi>1$ are possible due to  the softness of the repulsive force (see SI Appendix for simulation parameters).
 \label{ens} 
}
\end{figure}

\begin{figure}[t]
\centering
 \includegraphics[width= 0.97 \columnwidth ]{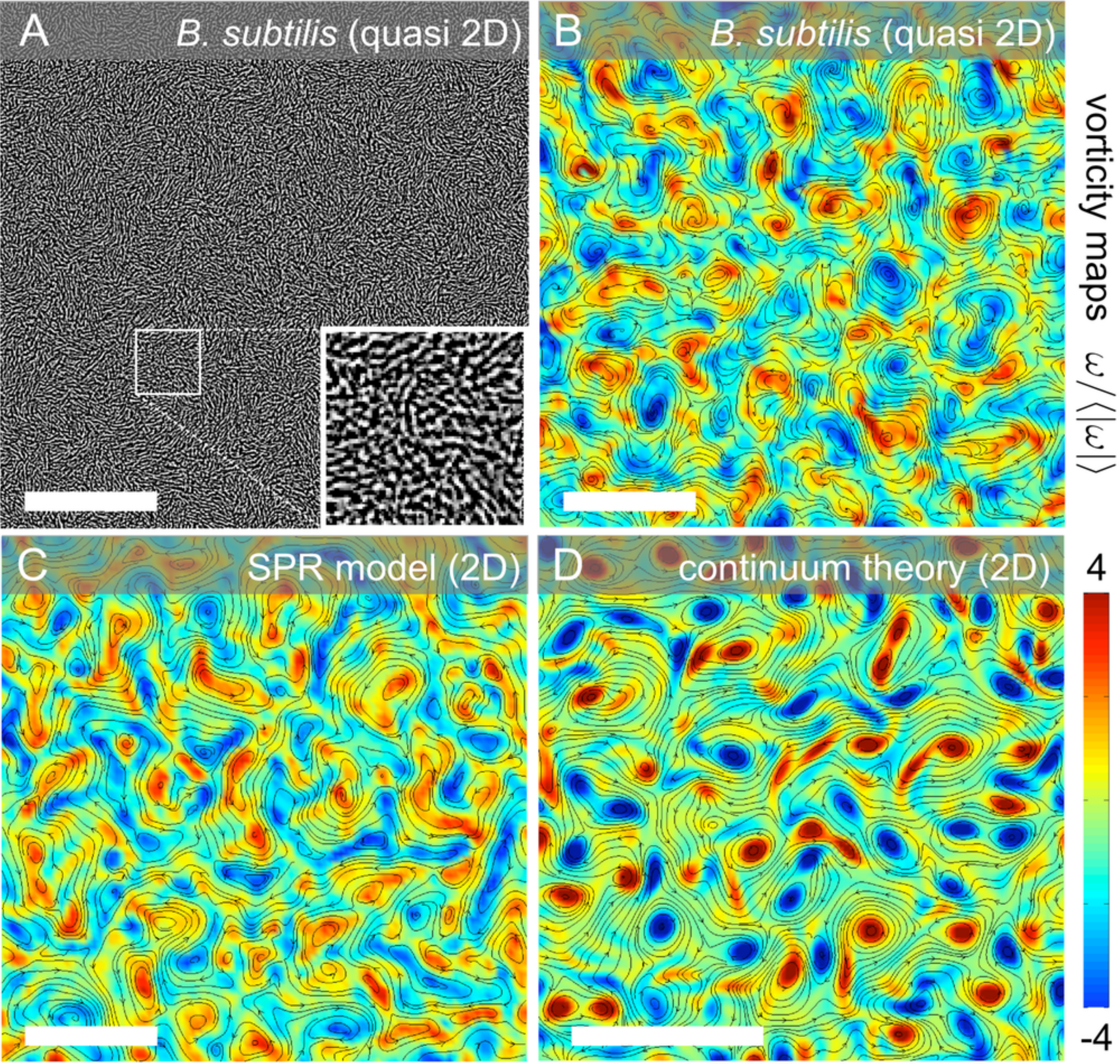}
\caption{Experimental snapshot (A) of a highly concentrated, homogeneous quasi-2D  bacterial suspension (see also Movie S07 and Fig. S8). Flow streamlines ${\boldsymbol v} (t,{\boldsymbol r} )$  and vorticity fields $\omega(t,{\boldsymbol r} )$   in the turbulent regime, as obtained from (B) quasi-2D bacteria experiments,  (C) simulations of the  deterministic SPR model ($a = 5$, $\phi = 0.84$), and (D) continuum theory.  The range of the simulation data in (D) was adapted to the experimental  field of view (217 $\mu$m $\times$ 217 $\mu$m)  by matching the typical vortex size  (scale bars $50 \mu$m). Simulation parameters are summarized in the SI Appendix.
 \label{turbsnap}}
\end{figure}

 A typical turbulent flow state as found in the simulations, and the associated (pseudo-scalar)  2D vorticity field $\omega=\partial_x v_y-\partial_y v_x$,  are shown in Fig.~\ref{turbsnap}.  
The mean local flow field ${\boldsymbol v} (t,{\boldsymbol r} )$ at time $t$ and position~$\boldsymbol
r$  was constructed by binning and averaging individual particle velocities, using a spatial resolution similar to that in our experiments (SI Appendix). 
To characterize the emergence of homogeneous turbulence in the SPR model in terms of particle geometry~$a$  and effective volume fraction~$\phi$, we quantify the vortical energy through the  enstrophy~\cite{2000Danilov,2002Kellay,2004Frisch}  per unit area, $
 {\Omega}  = \frac{1}{2} \,
 \overline{\left \langle    \left |\omega(t,{\boldsymbol r} )\right|^{2} \right \rangle},
 $
where brackets $\langle\,\cdot\,\rangle$ indicate spatial averages and overbars denote time averages.
 For slender rods ($a \geq 3$) the mean enstrophy $\Omega$ exhibits a maximum when plotted versus the volume fraction~$\phi$ (Fig.~\ref{ens}B). 
This maximum coincides approximately with the transition from the bionematic to the turbulent phase; in a bacterial suspension, it corresponds to the optimal concentration for fluid mixing. Typical aspect ratios of bacterial cell bodies in our experiments lie in the range $2\lesssim a\lesssim 10$ (mean $6.3\pm 1.2$; see SI Appendix, Fig.~S7). Hence, homogeneous bacterial turbulence should be observable in 2D for a broad range of filling fractions.

\subsection{Experiments} 

We test the T-phase  of the SPR model against experimental observations of \textit{B.~subtilis} at high filling fractions ($\phi\gtrsim 50\%$, see Materials and Methods). In contrast to recent investigations of bacterial dynamics in 2D free-standing films~\cite{2007SoEtAl},  on 2D surfaces~\cite{Swinney_bactclust,2012Chen_PRL}  and open 3D bulk suspensions~\cite{2004DoEtAl,2011Japan,2011Cisneros_PRE,2007Cisneros}, bacteria were  confined in closed microfluidic chambers to minimize oxygen gradients  that may cause anisotropic streaming of the oxytactic ~\textit{B.~subtilis} bacteria~\cite{2011KochSub}.  To study the effects of dimensionality and boundary conditions, experiments were performed with two different set-ups:  quasi-2D microfluidic chambers  with a vertical height $H$ less or equal to the individual body length of~\textit{B. subtilis} ($\sim 5 \,\mu$m), and  3D chambers with $H\approx 80\,\mu$m (Fig.~S6, S8 and Movies S07-S10).  To focus on the collective dynamics of the microorganisms  rather than the solvent flow~\cite{2011Japan,2010DuPuZaYe}, we determined the mean local motion of  \textit{B.~subtilis} directly using particle imaging velocimetry (PIV; see also SI Appendix). A typical snapshot from a quasi-2D experiment is shown in Fig.~\ref{turbsnap}A. As evident from the inset,  local density fluctuations which are important in the swarming/flocking regime~\cite{2007NaRaMe,2012Chen_PRL}  become suppressed at very high filling fractions (SI Appendix, Fig. S5). The corresponding flow  fields (Fig.~\ref{turbsnap}B and SI Appendix, Fig. S8) were used for the statistical analysis presented below.

\subsection{Continuum theory}

The analytical understanding of turbulence phenomena hinges on the availability of simple, yet sufficiently accurate continuum models~\cite{2004Frisch}. Considerable efforts have been made  to construct effective  field theories for active systems~\cite{2005ToTuRa,2010Ramaswamy,1998TonerTu_PRE,2008Wolgemuth,2009BaMa_PNAS,2008BaMa,2010Pedley,2002Ra,2008SaintillanShelley} but most of them have yet to be tested quantitatively against experiments.  Many continuum models distinguish solvent velocity, bacterial velocity and/or orientational order parameter fields,  resulting in a prohibitively  large number of phenomenological parameters and making comparison with experiments very difficult. Aiming  to identify a minimal hydrodynamic  model of self-sustained meso-scale turbulence,  we study a simplified continuum theory for incompressible active fluids, by focussing solely on the experimentally accessible velocity field $\boldsymbol v(t,\boldsymbol r)$. By construction, the theory will not be applicable to regimes where density fluctuations are large (e.g., swarming or flocking), but it can provide a useful basis for quantitative comparisons with  particle simulations  and experiments at high concentrations.
\par
We next summarize the model equations; a detailed motivation is given in the SI Appendix. Since our experiments suggest that density fluctuations are negligible (Fig. 2A) we postulate incompressibility,  $\nabla \cdot \boldsymbol v=0$. The dynamics of $\boldsymbol v$ is governed by an incompressible Toner-Tu equation~\cite{2005ToTuRa,2010Ramaswamy,1998TonerTu_PRE}, supplemented with a Swift-Hohenberg-type fourth-order term~\cite{1977SwiftHohenberg}, 
 \begin{eqnarray}
 (\partial_t + \lambda_0 \boldsymbol v\cdot \nabla) \boldsymbol v
 &=&\notag
-\nabla p  +\lambda_1 \nabla \boldsymbol v^2 - (\alpha+\beta |\boldsymbol v|^2)\boldsymbol v + 
\\&&
\Gamma_0 \nabla^2 \boldsymbol v -\Gamma_2(\nabla^2)^2    \boldsymbol v,
\label{e:conti-b}
 \end{eqnarray}
where $p$ denotes pressure,  and general hydrodynamic considerations~\cite{2010Pedley} suggest that $\lambda_0>1, \lambda_1>0$ for pusher-swimmers  like \emph{B. subtilis} (see SI Appendix). The $(\alpha, \beta)$-terms in Eq.~\eqref{e:conti-b} correspond to a quartic Landau-type velocity potential~\cite{2005ToTuRa,2010Ramaswamy,1998TonerTu_PRE}. For $\alpha>0$ and $\beta>0$, the fluid is damped to a globally disordered state with $\boldsymbol v=0$, whereas for $\alpha<0$ a global polar ordering is induced.  However, such global polar ordering is not observed in suspensions of swimming bacteria, suggesting that other instability mechanisms prevail~\cite{2002Ra}.  A detailed stability analysis  (SI Appendix)  of Eq.~\eqref{e:conti-b} implies that the Swift-Hohenberg-type $(\Gamma_0,\Gamma_2)$-terms provide the simplest  generic description of self-sustained meso-scale turbulence in incompressible active flow: For $\Gamma_0<0$ and $\Gamma_2>0$, the model exhibits a range of unstable modes, resulting in turbulent states as shown in Fig.~\ref{turbsnap}D. Intuitively,  the $(\Gamma_0,\Gamma_2)$-terms describe intermediate-range interactions, and their role  in Fourier-space is similar to that of the Landau-potential in velocity space (SI Appendix). We therefore expect that Eq.~\eqref{e:conti-b} describes a wide class of quasi-incompressible active fluids. To compare the continuum model with experiments and SPR simulations, we next study traditional turbulence measures.

\subsection{Velocity Structure Functions} 

Building on  Kolmogorov's seminal work~\cite{1941Kolmogorov_1},  a large part of the classical turbulence literature~\cite{2004Frisch,2002Kellay,2002Gotoh_PhysFluids,1997Noullez_JFM,1997Camussi,1999LewisSwinney_PRE,1984Anselmet_JFM} focuses on identifying the distribution of the flow velocity increments \mbox{$\delta \boldsymbol v(t, \boldsymbol r, \boldsymbol R)=\boldsymbol v(t,\boldsymbol r+\boldsymbol R)- \boldsymbol v(t,\boldsymbol r)$}. Their statistics is commonly characterized in terms of 
the longitudinal and transverse projections,  \mbox{$\delta v_{||}=\hat{\boldsymbol R}\cdot \delta \boldsymbol v$} and \mbox{$\delta v_{\perp}=\hat{\boldsymbol T}\cdot \delta \boldsymbol v$}, where $\hat{\boldsymbol T}=(\epsilon_{ij} \hat{R}_j)$ denotes a unit vector perpendicular to the unit shift vector $\hat{\boldsymbol R}=\boldsymbol R/|\boldsymbol R|$. The separation-dependent statistical moments of $\delta v_{||}$ and $\delta v_{\perp}$ define the longitudinal and transverse velocity structure functions 
\begin{equation} 
\label{e:k-structure} 
S^n_{||,\perp}({\boldsymbol R}) :=
\left \langle\bigl(  \delta v_{||,\perp} \bigr)^{n}
\right \rangle
,\qquad
n=1,2,\ldots\;.
\end{equation} 
These functions have been intensely studied in turbulent high-Re fluids~\cite{2004Frisch,2002Kellay,2000Danilov,1984Anselmet_JFM}, but are unknown  for active flow. For isotropic steady-state turbulence, spatial averages $\langle\, \cdot\, \rangle$ as in  Eq.~\eqref{e:k-structure} become time-independent, and the moments  $S^n_{||,\perp}$ reduce to functions of the distance $R=|\boldsymbol R|$.

%%%%%%%%%%%%%%
\begin{figure*}
\centering
 \includegraphics[clip=,width= 1.8  \columnwidth ]{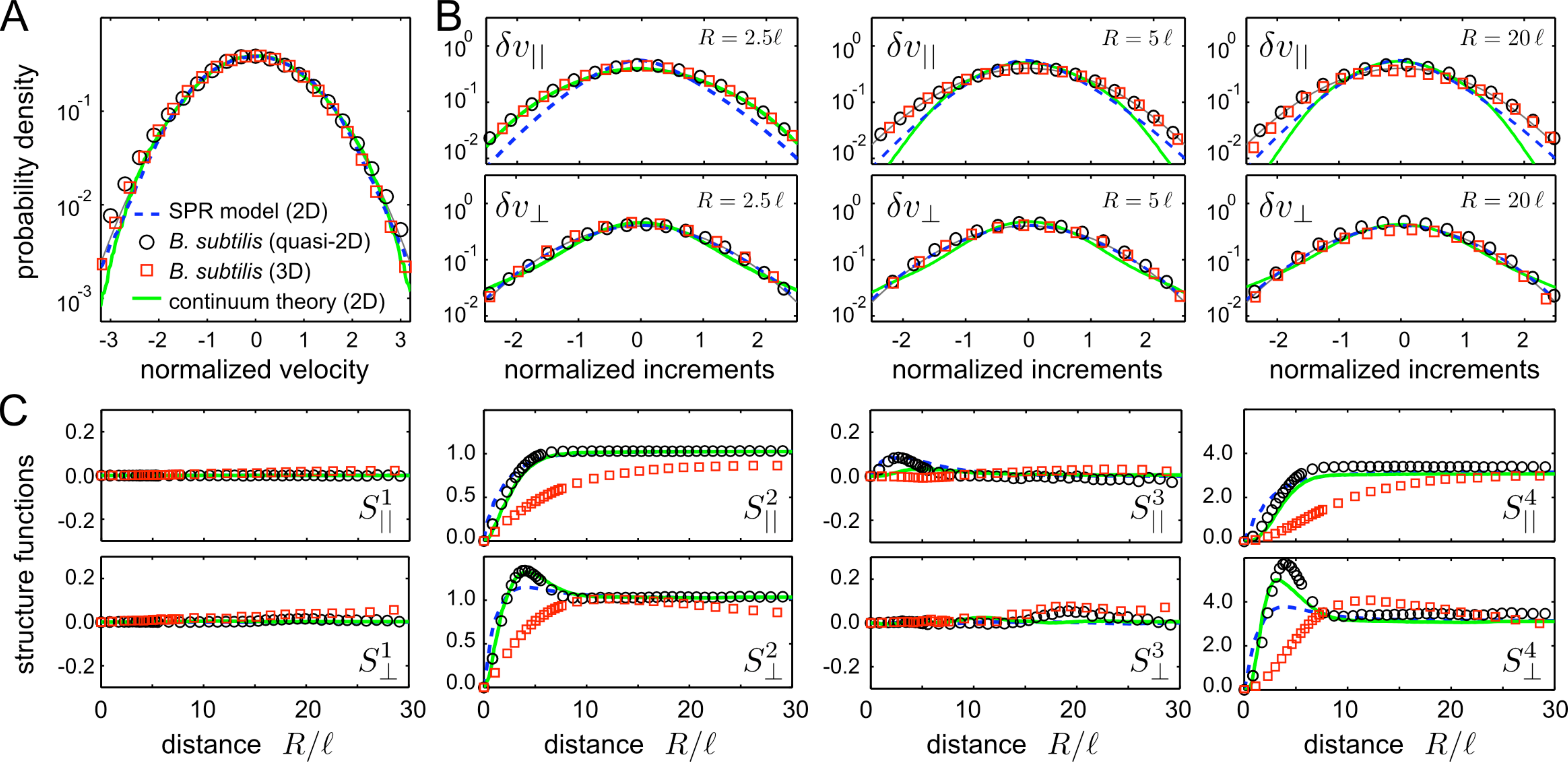}
\caption{
Velocity statistics of self-sustained turbulent phases in active suspensions. (A)~The marginal distributions of the normalized Cartesian velocity components $[v_{i}-\langle v_{i}\rangle] / [\langle v_i^2\rangle -\langle v_{i}\rangle^2]^{1/2}_{i=x,y}$  are approximately Gaussian (thin grey line) for experiments, SPR model and continuum theory. (B)~The distributions of the longitudinal and transverse velocity increments  $\delta v_{||,\perp}$, normalized by their first and second moments $S^{1,2}_{||,\perp}$ are shown for three different separations $R$.  (C)~Longitudinal and transverse velocity structure functions~$S^n_{||,\perp}$ normalized by $\overline{\langle v^2\rangle}^{n/2}$. The maxima of the even transverse structure functions~$S^{2k}_{\perp}$ reflect  the typical vortex size $R_v$  which is significantly larger in the 3D experiments. Experimental and theoretical data points are spatio-temporal averages over two orthogonal directions in (A) and (B), and all directions in  (C), yielding a typical sample size  $>10^6$ per plotted data point in (C).   Histograms and structure functions for quasi-2D (3D) curves were obtained by combining PIV data from two (fifteen) movies, respectively, representing an average over 2x1000 (15x300) frames.  Simulation parameters are identical to those in Fig.~\ref{turbsnap} and summarized in the SI Appendix. Errorbars are smaller than symbols.
 \label{fig:SF}}
\end{figure*}
%%%%%%%%%%%%%%%

\par %%% SPR structure functions: technical
Velocity distributions, increment distributions and structure functions for our numerical and experimental data are summarized in Fig.~\ref{fig:SF}. For the SPR model, the velocity statistics can be calculated either from the raw particle data or from pre-binned flow field data. The two methods produce similar results, and Fig.~\ref{fig:SF} shows averages based on individual particle velocities. Generally, we find that both the 2D SPR model and the 2D continuum simulations are capable of reproducing the  experimentally measured quasi-2D flow histograms  (Fig.~\ref{fig:SF}A,B) and structure functions  (Fig.~\ref{fig:SF}C). The maxima of the even transverse structure $S^{2n}_\perp$ signal a typical vortex size $R_v$, which is substantially larger in 3D  bulk flow than in quasi-2D bacterial  flow. Unlike their counterparts in high-Re Navier-Stokes flow~\cite{2004Frisch,2002Kellay},  the structure functions of active turbulence exhibit only a small region of  power law growth for $\ell\lesssim R\ll R_v$ and flatten at larger distances (Fig.~\ref{fig:SF}C).

%%%%%%%%%%%%
\subsection{Velocity Correlations and Flow Spectra} 
\begin{figure}[b]
\centering
 \includegraphics[clip=,width=  \columnwidth ]{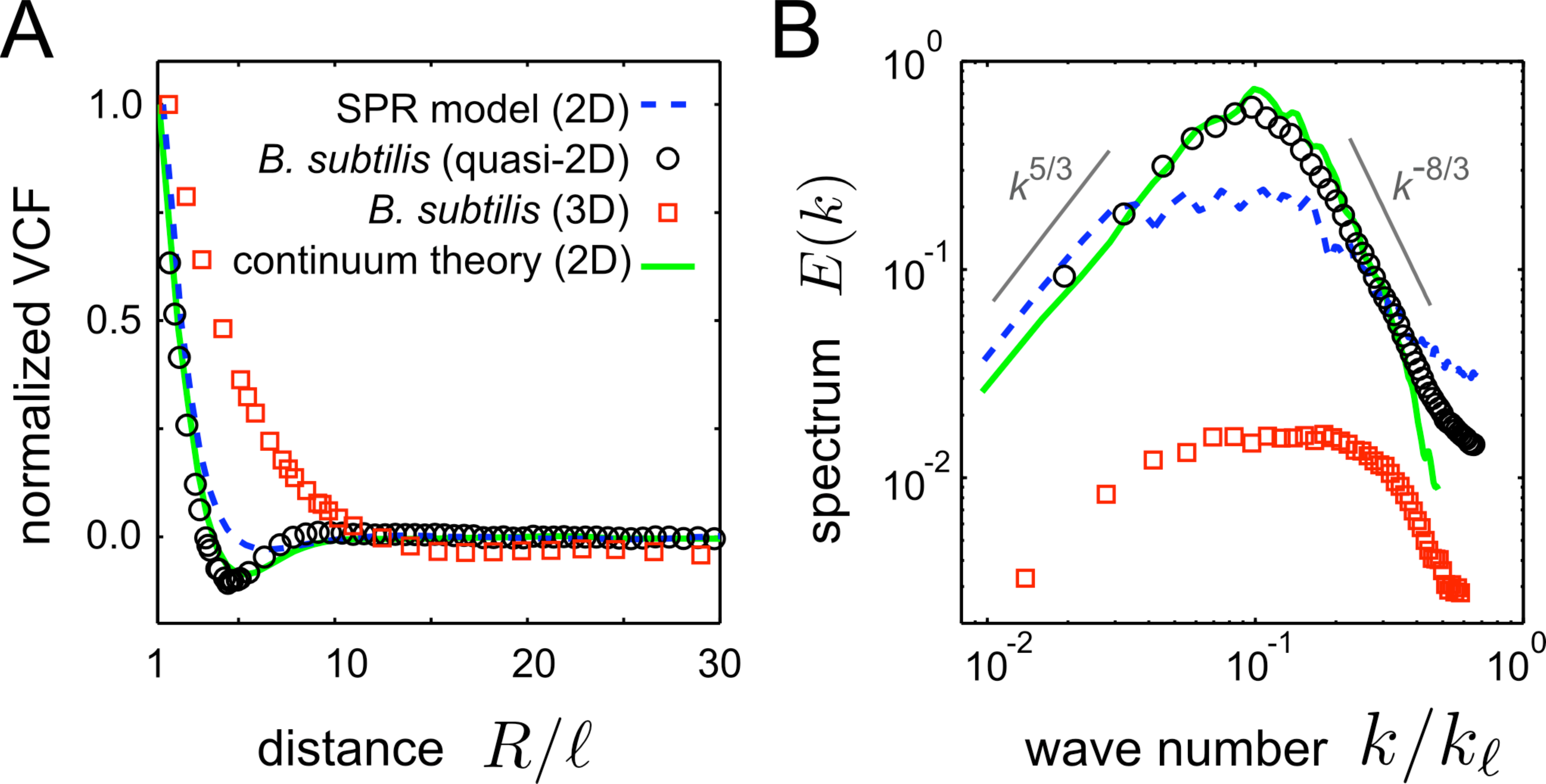}
\caption{Equal-time velocity correlation functions (VCFs), normalized to unity at $R=\ell$,  and flow spectra for the  2D SPR model  ($a=5,\,\phi=0.84$),  {\it B. subtilis} experiments, and 2D continuum theory based on the same data as in Fig.~\ref{fig:SF}. (A)  The minima of the VCFs reflect the characteristic vortex size $R_v$~\cite{2009Swinney_EPL}.  Data points present averages over all directions and time steps to maximize sample size. (B)  For bulk turbulence (red squares) the 3D spectrum $E_3(k)$ is plotted ($k_\ell=2\pi/\ell$), the other curves show 2D spectra $E_2(k)$.  Spectra for the 2D continuum theory and quasi-2D experimental data are in good agreement; those of the 2D SPR model and the 3D bacterial data show similar asymptotic scaling but exhibit an intermediate plateau region    
(spectra multiplied by constants for better visibility and comparison). 
 \label{f:spectrum} }
\end{figure}

%%%%%%%%%%%%
The energy spectrum $E(k)$, formally defined by $\langle \boldsymbol v^2 \rangle=2\int_0^\infty E(k) dk$, reflects the accumulation of kinetic energy over different length scales. By virtue of the Wiener-Khinchine theorem~\cite{2004Frisch}, $E(k)$ can be estimated by Fourier-transformation of the equal-time two-point velocity correlation function,  yielding in $d$ dimensions
\begin{equation} \label{e:spec}
E_d(k) =  \frac{k^{d-1}}{C_d} \int d^d R\; e^{-i {\boldsymbol k} \cdot \boldsymbol R } \;
\langle {\boldsymbol v} (t,\boldsymbol r) \cdot {\boldsymbol v} (t,\boldsymbol r+\boldsymbol R)\rangle,
\end{equation}  
where $C_2=2\pi$ and $C_3=4\pi$. Normalized velocity correlation functions $\overline{\langle {\boldsymbol v} (t,\boldsymbol r) \cdot {\boldsymbol v} (t,\boldsymbol r+\boldsymbol R)\rangle}$ and spectra $E_d(k)$ for our data are summarized in Fig.~\ref{f:spectrum}.
The crossover from positive to negative correlations indicates again the typical vortex size $R_v$, in agreement with Fig.~\ref{fig:SF}C and previous findings for open 3D bulk systems~\cite{2004DoEtAl,2007Cisneros}.  

\par
In bacterial suspensions,  the microorganisms inject kinetic energy on small scales $R\sim\ell$, setting the upper bound $k_\ell=2\pi/\ell$ for the spectral range of the bacterial fluid. For both experiments and simulations, we observe turbulent vortices on scales $R>\ell$, which formally  correspond to the energy-inertial range $k< k_\ell$ in classical 2D turbulence~\cite{2002Kellay,2000Danilov}.  Our experimental  and numerical data suggest asymptotic  power law scaling regimes for small and large $k$-values (see Fig.~\ref{f:spectrum}B), but the power-law exponents differ from the characteristic $ k^{-5/3}$-decay of 2D Kolmogorov-Kraichnan turbulence~\cite{1980Kraichnan}; see discussion below. The spectra for the 2D continuum model and the quasi-2D bacteria experiments are in good agreement, both showing large-$k$ scaling with approximately $E(k)\sim k^{-8/3}$ and small-$k$ scaling with roughly $E(k)\sim k^{+5/3}$. The asymptotic  spectra for the 2D SPR model and the 3D experimental data look qualitatively similar, but do also exhibit an intermediate plateau region which indicates that kinetic energy is more evenly distributed over a range of scales.

%%% continuum model data from run 453 %%%%
%%%%%%%%%%%%%%%

%%%%%%%%%%%%%%%%%%%%
%\newpage
\section{Discussion and Conclusions}
%%%%%%%%%%%%%%%%%%%%

\subsection{SPR Model vs. Experiment}
The deterministic SPR model provides a simplified description of the bacterial dynamics, as it neglects not only elastic properties of flagella and cell body, but also hydrodynamic interactions and orientational fluctuations due to intrinsic swimming variability and thermal effects~\cite{2011DrescherEtAl,2009Hoefling}. Notwithstanding, at high concentrations, such a minimal model reproduces remarkably well the flow velocity distributions and the structure functions from  our quasi-2D~\textit{B. subtilis} experiments and the 2D continuum simulations~(Fig.~\ref{fig:SF}). This implies that hydrodynamic interactions \textit{per se} are not required for the formation of self-sustained turbulence in dense suspensions -- self-propulsion, a rod-like shape and volume exclusion interactions are sufficient (this raises the question whether the optimization of collective behavior may have been a factor in the evolution of bacterial shapes). However, to achieve a better quantitative agreement, particle-based future studies should focus on more realistic models that account for hydrodynamic near-field interactions and intrinsic randomness in bacterial swimming~\cite{2011DrescherEtAl}. The experimental results presented above provide a benchmark for evaluating such microscopic models~\cite{2009Graham}. 

\subsection{Continuum Model and \lq Universality\rq}
The good agreement of the structure functions, spatial and temporal flow correlations (see also Fig. S9), and spectra from the 2D continuum theory with those from  the quasi-2D experiments suggests that this theory could be a viable model for dense suspensions. Since the instability mechanism in the continuum theory arises from a generic  small-wave number expansion in Fourier-space (see SI Appendix),  that is analogous to the Landau-expansion in order-parameter space for second-order phase-transitions, we expect that the model applies to a wide range of quasi-incompressible active fluids. This would imply that meso-scale turbulent structures in these systems share \lq universal\rq\space long-wave length characteristics.  We note that the theory as formulated in Eq.~\eqref{e:conti-b} only accounts for leading terms up to fourth-order  and, therefore, becomes inaccurate for large velocities and wave numbers (see tails in Figs.~3A,B and 4B). Nevertheless, this continuum model appears to capture the main statistical and dynamical features of the experimental data.  Important future challenges include the  analytical prediction of active flow spectra from Eq.~\eqref{e:conti-b}, detailed  numerical studies of 3D bacterial bulk flows, and  comparisons of our experimental and numerical data with $\boldsymbol Q$-tensor models and other multi-order parameter theories~\cite{2011KochSub,2010Ramaswamy,2008Wolgemuth,2009BaMa_PNAS,2008BaMa}.

\subsection{Dimensionality, Boundaries and Hydrodynamic Interactions}
The quasi-2D experiments allow us to compare with 2D simulations that come close to experimental system sizes. Free-standing thin films~\cite{2007SoEtAl} and bacterial mono-layers on open surfaces~\cite{Swinney_bactclust,2012Chen_PRL}, which may be more prone to intrinsic instabilities and external fluctuations, provide an alternative, but non-equivalent realization of a 2D bacterial fluid. The crucial difference between freestanding 2D films and our closed quasi-2D set-up is that the presence of no-slip boundaries in our experiments suppresses hydrodynamic long-range interactions between bacteria due to cancellation effects from the hydrodynamic images: An isolated  dipole-like swimmer (as {\it E. coli}~\cite{2011DrescherEtAl} and, most likely, {\it B. subtilis}) creates a stroke-averaged far-field flow that decays as $\sim 1/r^2$ with distance $r$ in a 3D fluid. When the same swimmer moves parallel to a nearby solid surface in an otherwise semi-infinite fluid, the flow components parallel to the boundary decay faster $\sim1/r^4$~\cite{2011DrescherEtAl}. If, however, the swimmer is closely confined between two parallel no-slip walls, as in our quasi-2D experiments with $H\sim 4\;\mu$m, then the flow field becomes exponentially damped at  distances $|r|\gg H$~\cite{1976LironMochon}.  By contrast, in free-standing 2D films the flow field generated by an isolated microorganism has a much longer range~$\sim 1/r$~\cite{2010Gollub_PRL,2011Gollub_PNAS}, suggesting that hydrodynamic interactions could play a more important role for collective behavior in these systems~\cite{2007SoEtAl}. The fact that the typical vortex size in 3D  is larger 
than in quasi-2D  could indicate stronger short-to-intermediate-distance hydrodynamic coupling in 3D bulk flow;  it would therefore be interesting to perform a similar analysis 
for thin-film data~\cite{2007SoEtAl}. Generally, however, we expect  hydrodynamic far-field interactions to be less important for the dynamics in very dense suspensions due to mutual hydrodynamic screening~\cite{1983MuEd_Screening} and the small magnitude of bacterial flows fields ~\cite{2011DrescherEtAl}, but they could act as a  destabilizing noise~\cite{2008SaintillanShelley, 2011Aranson_PRE}.

\subsection{Low-Re vs. High-Re Turbulence}
Conventional high-Re turbulence arises from energy input on large scales (e.g., stirring or shearing). In 3D flow the injected energy is redistributed to smaller scales via an energy-inertial downward cascade with $E_{3}\sim k^{-5/3}$~\cite{2004Frisch}. In 2D films, due to the suppression of vortex stretching~\cite{2000Danilov,2002Kellay}, there can be both an energy-inertial upward cascade with $E_{2}\sim k^{-5/3}$ and an enstrophy-transfer downward cascade with $E_{2}\sim k^{-3}$~\cite{1980Kraichnan}. 
Remarkably,  viscoelastic polymer solutions can exhibit turbulent features (e.g., spectral power law scaling) at Reynolds numbers as low as $10^{-3}$, facilitated by a slow nonlinear response to external shear due to long intrinsic relaxation times of the polymers~\cite{2000Groisman,2010GoddardHess}.  Our simulations and experiments suggest asymptotic spectral power law decays towards the bacterial energy injection scale $k_\ell=2\pi/\ell$ that resemble the energy-inertial regime of classical turbulence but, due to viscous damping by the low-Re solvent, extend over a smaller range of length scales (roughly up to $10\ell$). The latter fact is reminiscent of viscoelastic turbulence~\cite{2000Groisman}, although the underlying physical mechanisms are very different.
\par
In conclusion, bacterial or, more generally, self-sustained \lq active turbulence\rq,~shares some qualitative characteristics with classical turbulence on small scales while differing on larger scales.   
Our detailed statistical analysis shows that, as with inertial turbulence, a complete quantitative understanding of turbulent behavior in active systems  poses a challenging task. The combined  experimental, theoretical and numerical results presented here may provide both qualitative and quantitative guidance for future studies that  aim at identifying the basic principles of dynamical self-organization in living fluids. 

\begin{acknowledgments}
The authors are grateful to Gareth Alexander, Thomas Angelini, Igor Aronson,  Markus B\"ar, Howard Berg, Colm Connaughton, Sujoy Ganguly, Siegfried Hess, Vasily Kantsler, John Lister, Peter Lu, Timothy Pedley, Adriana Pesci and David Weitz for very helpful discussions. S.H. acknowledges financial support from the Deutsche Forschungsgemeinschaft (DFG), Grant HE5995/1-1. This work was also supported by DFG via SFB TR6 (section D3), EPSRC and ERC.
\end{acknowledgments}

%%%%%%%%%%%%%%%%%%%%%%%%%%%%
%%%%%%%%%%%%%%%%%%%%%%%%%%%%
%\pagebreak
\begin{materials}
\textit{B. subtilis} cells (wild type strain 168) were streaked from a -80$^{\circ}$C stock onto an LB medium plate containing 1.5\% agar. The plates were incubated at 37$^{\circ}$C for 12 h. A single colony from the plates was used to inoculate an overnight culture in Terrific Broth (Sigma), which was then back-diluted 1:200 into 50 mL of fresh tryptone broth, and grown at 37$^{\circ}$ C on a shaker to mid-log phase. The culture was then concentrated 400$\times$ by centrifugation at 4000$\times g$ for 3 min, and the pellet was resuspended by gentle vortexing, to not shear off the flagella. The concentrated culture was loaded into a polydimethylsiloxane (PDMS) microfluidic device, which was then sealed to reduce background fluid motion. The microfluidic device consisted of cylindrical measurement chambers (radius 100 $\mu$m, height 4 $\mu$m for quasi-2D measurements, and radius 750 $\mu$m, height 80 $\mu$m for 3D measurements). The samples were imaged in bright field with a 40$\times$/NA 1.4 oil immersion objective on a Nikon TI-E microscope. Images were acquired at 40 fps in 2D (camera: Pike, Allied Vision Technologies), and 100 fps and 200 fps in 3D (camera: Phantom v9.1, Vision Research). Compared with measurements in quasi-2D chambers at the  same frame rate, the vertical superposition of bacteria leads to a reduced image quality in 3D  samples; we therefore recorded the flow in 3D suspensions at a higher frame rate.  For the 3D measurements, we imaged at the bottom and in the middle of the chamber, while for the quasi-2D measurements, we imaged in the middle of the chamber.  
A detailed description of the theoretical models and numerical methods is 
given in the SI Appendix. Raw data and additional experimental movies can be downloaded from: http://damtp.cam.ac.uk/user/gold/datarequests.html
\end{materials}

%%%%%%%%%%%%

\end{article}

\end{document}